\documentclass[12pt]{article}
\usepackage[totalheight = 23cm, totalwidth = 17cm]{geometry}

\newcommand{\be}{\begin{equation}}
\newcommand{\ee}{\end{equation}}
\newcommand{\bea}{\begin{eqnarray}}
\newcommand{\eea}{\end{eqnarray}}

\newcommand{\nn}{\nonumber\\}

\usepackage{graphicx}

\begin{document}

\begin{center}

{\bf{\Large Dark Energy as an off-shell Tachyon Background in four-dimensional Strings}}

\vspace{1cm}

J. Alexandre\footnote{jean.alexandre@kcl.ac.uk} and
N. E. Mavromatos\footnote{nikolaos.mavromatos@kcl.ac.uk}\\
Department of Physics, Kings's College London, WC2R 2LS, U.K.

\vspace{0.5cm}

{\bf Abstact}

\end{center}

We consider a time-dependent bosonic string in graviton, dilaton and tachyon backgrounds, for which it was already shown that conformal invariance is respected to all orders in $\alpha'$ and in any space-time dimension.
Assuming that the tachyon is off-shell, we show in this note that the specific time-homogeneity of the corresponding space time effective action leads to a power-law expanding Universe, that can be interpreted as dark energy.
We arrive at this result without requiring the knowledge of the structure of the string effective action,
which is not known to all orders in $\alpha '$. Moreover, in our approach,
the background configurations are consistent in a four-dimensional space time, without any need for extra dimensions.

\vspace{1cm}

\section{Introduction and Summary}

One of the most important experimental discoveries of the past decade is the fact that at the current era our Universe is accelerating. The simplest explanation for this phenomenon is that the energy budget of our Universe is populated by an unknown form of energy, possibly quantum in origin, termed Dark Energy. In its simplest version, Dark Energy is just a positive cosmological constant, and the associated Gravitational system is nothing other than a de Sitter Universe.  In the context of a homogeneous and isotropic Friedmann-Robertson-Walker (FRW) Cosmology, the de-Sitter model, in its simplest version,  can match a plethora of astrophysical observations, ranging from galactic and super-galactic cluster dynamics to Cosmic Microwave Background temperature fluctuations, provided there is a sufficient amount of cold dark matter. Such a model, termed as the $\Lambda$CDM model (Cold Dark Matter with a Cosmological constant $\Lambda > 0$), is as successful for Modern Cosmology as the Standard Model of Particle Physics, at present,
but is not a consistent background of perturbative string theory \cite{hellerman}

In a first quantized string framework, in which the dynamics of strings propagating in certain non-trivial space-time backgrounds is described by appropriate world-sheet $\sigma$-models, the r\^ole of local conformal invariance on the world-sheet of the string is
essential to leave space time Physics unaffected.
In \cite{AEM1} we have addressed this latter issue by studying the exact 
$\alpha'$-dependence of the quantized world-sheet $\sigma$-model,
and we found conformal invariant backgrounds, which are characterized by a theory independent of the value of $\alpha'$, and therefore 
of the amplitude of quantum fluctuations in the world-sheet theory. These properties have been demonstrated 
explicitly for bosonic strings propagating in \emph{time-dependent} graviton, dilaton, antisymmetric tensor \cite{aet} and 
tachyon backgrounds \cite{AKM}. An important point is that such backgrounds were consistent with conformal invariance in four space-time dimensions~\cite{AM}, among other critical values.

Local conformal invariance on the world sheet of the string can be guaranteed for these configurations
only by means of appropriate field redefinitions. The homogeneity in time of the various terms
in the expressions of the world-sheet Weyl-anomaly coefficients guarantees the existence
of a \emph{given renormalization scheme} in which the $\beta$-functions vanish.
In this sense, the corresponding backgrounds are interpreted as classical on-shell solutions of equations of motion obtained from an effective action, whose form cannot be known explicitly, due to the complicated structure of the various terms
(infinite in order of derivatives). Thus, the scheme in which the various $\beta$-functions are expected to vanish
is only formally proven to exist. 

We should stress at this point that the string configurations found in \cite{AEM1,aet,AKM} as solutions 
leading to $\alpha'$-independent theories {\it do not} satisfy target-space Einstein's equations:
conformal invariance of these configurations is not obtained by the vanishing of one-loop Weyl anomaly $\beta$-functions. 
Instead, conformal invariance
arises from the homogeneity of the terms of these $\beta$-functions in the target time $X^0$,
to all orders in $\alpha'$. For this reason, the corresponding space-time effective action, although not
known explicitly, nevertheless contains terms homogeneous in the target time. It should be notice that
these exact solutions in $\alpha '$ could not have been found in generic string theory backgrounds, but
they are specific to cosmological backgrounds that are isotropic and homogeneous (in the cosmological
sense~\footnote{The reader should not confuse here the different uses of the world ``homogeneity'': in one case,
homogeneity is used to denote the mathematical property of the dependence of the various terms of the $\beta$-functions
on a single power of $X^0$, while in the other (cosmological) sense, the terminology ``homogeneous'' is used to denote
the fact that there is no preferred point in the target-space Universe.}), and hence depending only on $X^0$.

In \cite{AKM} we have included tachyonic backgrounds in this setting, for which the 
difference with other ones (dilaton, graviton and antisymmetric tensor) lies on the fact that there are 
terms in the tachyon $\beta$-functions that are inhomogeneous in the time $X^0$.
We have argued, however, that by performing appropriate field redefinitions, 
order by order in $\alpha '$, we can be lead to a scheme in which these inhomogeneous terms are absent.

In this article we demonstrate that, with a dilaton and graviton background, 
conformal invariance obtained from $\beta$-functions which are homogeneous in time
leads to a target-space Minkowski Universe, which we consider as ``{\it vacuum state}'' of our theory.
We then consider the case of an \emph{off-shell} tachyon background,
in which the inhomogeneous terms of the corresponding tachyon $\beta$-function are present,
and as a result the $\beta$-function is non-vanishing,
despite the fact that the tachyon configuration doesn't spoil the $\alpha'$-independence of the world sheet quantum theory.
We will show that this leads to a configuration which
represents an ``\emph{excited state}'' of the Universe, and breaks the homogeneity (in time)
of the target-space effective action. Such configurations imply a power law expanding Universe,
which can be decelerating or accelerating Universe, depending on the range of parameters in the solution.
We interpret this as signal of dark energy, and the important point to notice is that we have arrived at such a conclusion without knowledge of the structure of the effective target-space action.

We note that in our context, dark energy cannot be seen as a cosmological constant, 
but rather as a infinite series of corrections to Einstein's equations, including all the $\alpha'$-loop orders, 
which are homogeneous (in time) to the Einstein term in the action. 
The precise form of these corrections is not known, and is not even unique, since higher orders in $\alpha'$ 
are not fixed but can be modified by appropriate field redefinitions \cite{metsaev}.

The structure of the article is the following: in the next section we review  graviton and dilaton backgrounds, which satisfy conformal invariance conditions to all orders in $\alpha '$, as a result of the homogeneity
in time $X^0$ of the various terms in the corresponding world-sheet renormalization group $\beta$-functions.
In section \ref{sec3} we discuss the generic construction of the target-space effective action 
from the world-sheet partition function, and demonstrate how our dilaton and graviton configurations 
lead to a Minkowski space time,  
without explicit knowledge of the structure of the effective action.
In section \ref{sec4} we include off-shell tachyons, in the sense of inhomogeneous (in time $X^0$) terms in the tachyon
$\beta$-function, not removed by the field redefinitions that have been used for the removal of the homogeneous terms.
We demonstrate how their inclusion leads to dark energy, again independently of the form of the target-space effective action.
Finally, conclusions are presented in section \ref{sec5}.

\vspace{0.5cm}

\section{Graviton and dilaton backgrounds: conformal invariance to all orders in $\alpha'$\label{sec2}}

We consider a bosonic string on a spherical world sheet, in graviton and dilaton time-dependent background, defined by the
two-dimensional bare action~\footnote{For brevity we shall omit, throughout this work, the antisymmetric tensor backgrounds. 
Their inclusion is straightforward, and does not present any conceptual difference in comparison to the graviton-dilaton 
system examined here.}
\be
S=\frac{1}{4\pi\alpha'}\int d2\xi\sqrt\gamma\left\{g_{\mu\nu}(X^0)\gamma^{ab}\partial_a X^\mu\partial_b X^\nu
+\alpha'R^{(2)}\phi(X^0)\right\}.
\ee
The quantization of this theory is based on the partition function $Z[g,\phi]$,
functional of the background $(g_{\mu\nu},\phi)$,
and parametrized by $\alpha'$. As was shown in \cite{AEM1}, it is possible to find configurations such that
the quantum theory is independent of the value of $\alpha'$, i.e. $\dot Z=0$, where a dot represents a
derivative with respect to $\alpha'$. This is what ``fixed point'' stands for in our context.
It was found in \cite{AEM1} that this configuration must necessarily satisfy $g_{\mu\nu}\propto\phi''\eta_{\mu\nu}$,
where a prime denotes a derivative with respect to $X^0$. A specific configuration which satisfies this
condition is
\be\label{config}
g_{\mu\nu}=\frac{\alpha'A}{(X^0)^2}\eta_{\mu\nu}~~~~~~~~~~\phi=\phi_0\ln\left(\frac{X^0}{\sqrt{\alpha'}}\right),
\ee
where $A$ and $\phi_0$ are dimensionless constants, and
leads to Weyl-invariance $\beta$-functions which are homogeneous in $X^0$, to all orders in $\alpha'$:
\be
\beta^g_{00}=\frac{\alpha'E_0}{(X^0)^2}\eta_{00}~~~~~~~~\beta^g_{ij}=\frac{\alpha'E_1}{(X^0)^2}\eta_{ij}
~~~~~~~~\beta^\phi=E_2,
\ee
where $E_0,E_1,E_2$ are dimensionless constants, independent of $\alpha'$.
This homogeneity in $X^0$ and therefore in $\alpha'$ is at the origin of the equivalence between conformal
invariance of the configuration (\ref{config}) and $\dot Z=0$.
Using field redefinitions, which for the configuration (\ref{config}) do not change the $X^0$-dependence,
it is then possible to cancel $\beta$-functions to all orders, leading to
a non-perturbative conformal invariance, for any space time dimension $D$. The metric appearing in the
configuration (\ref{config}) corresponds, in the string frame, to a de Sitter Universe, but in the Einstein frame,
the dilaton is
\be\label{dilaton}
\phi(t)=\frac{2-D}{2}\ln\left(\frac{t}{\sqrt{\alpha'}}\right) ,
\ee
and leads to a power-law expanding Universe, with scale factor
\be\label{scale}
a(t)\propto t^{1+\frac{D-2}{2\phi_0}}.
\ee
As a result, a Minkowski Universe is obtained if
\be\label{minkowski}
D-2+2\phi_0=0,
\ee
and it was shown in \cite{AM} that $D=4$ is a natural choice for the configuration (\ref{config}) to be consistent
with a Wick rotation in a Minkowski Universe. In fact, upon such analytic continuations, the $D=4$ case is the lowest order in a series of discrete solutions, defining new critical dimensions of the bosonic string in the non-trivial dilaton backgrounds.  For $D=4$ and $\phi_0=-1$ (required in order to
have a Minkowski Universe), we note that the string coupling $e^\phi$ is proportional to $1/t$ and allows
the perturbative nature of string loops (higher world-sheet topologies) for late times. In this sense, our solutions for a late-epoch in the string Universe's history, do not suffer from the well-known~\cite{gross} Borrel-non-re-summability of the string loop expansion.\\
Finally, we remind here that the dilaton (\ref{dilaton}) and scale factor (\ref{scale}) are {\it not} solutions of 
equations of motion obtained by varying a space time effective action, but are given by the homogeneity of $\beta$-functions,
to all orders in $\alpha'$.

\vspace{0.5cm}

\section{Space time effective action\label{sec3}}

It is known \cite{tseytlin1} that the space time effective action $\Sigma$, functional of the backgrounds, is given by
\be\label{sigmadef}
\Sigma[g,\phi]=-\left(\frac{dZ}{d\ln\epsilon}\right)_{\epsilon=1},
\ee
where $\epsilon$ is a dimensionless world sheet regulator, equal to the ratio of a short distance cut-off over the radius of the
spherical world sheet. Since $\dot Z=0$ for the configuration (\ref{config}),
the space time effective action is also independent of the value of $\alpha'$:
\be
\dot\Sigma=-\frac{d}{d\alpha'}\left(\frac{dZ}{d\ln\epsilon}\right)_{\epsilon=1}
=-\left(\frac{d\dot Z}{d\ln\epsilon}\right)_{\epsilon=1}=0,
\ee
where we note that $\epsilon$ and $\alpha'$ are independent parameters. The definition (\ref{sigmadef}) of the
effective action gives
\be
\Sigma=\frac{1}{\kappa_0^2}\int d^Dx\sqrt{|g|}e^{-2\phi}\left\{\frac{D-26}{6\alpha'}-
\frac{1}{2}R-\nabla^2\phi+\Delta\right\},
\ee
where $\Delta$ represents all the higher order terms in $\alpha'$.
For the specific configuration (\ref{config}),
\be\label{sigma}
\Sigma=\frac{A^{D/2}}{\kappa_0^2}(\alpha')^{\phi_0+D/2-1}\int\frac{d^Dx~\Theta}{(x^0)^{D+2\phi_0}},
\ee
where $\Theta$ is constant, independent of $x^0$ because of homogeneity, and therefore
also independent of $\alpha'$, since there is no other dimensionful
constant (besides the overall normalization constant $\kappa_0$).
For $\dot\Sigma=0$ to hold, we must therefore satisfy
\be
\phi_0+\frac{D}{2}-1=0,
\ee
which is exactly the condition (\ref{minkowski}) to obtain a Minkowski Universe.
As a consequence, conformal invariance of the configuration (\ref{config}), which is equivalent to
$\dot\Sigma=0$, leads to a Minkowski Universe, in any dimension. This defines our cosmological vacuum state,
and we now describe how to go away from this vacuum by considering a tachyonic background.

\vspace{0.5cm}

\section{Inclusion of (off-shell) Tachyons and Dark Energy\label{sec4}}

An extension of the work \cite{AEM1} and including tachyons was done in \cite{AKM}, where conformal invariance
can be satisfied in a similar way (using the $X^0$-homogeneity of the $\beta$-functions),
 with the configuration (\ref{config}) extended with the tachyon
\be\label{tachyon}
T=\tau_0\ln\left(\frac{X^0}{\sqrt{\alpha'}}\right),
\ee
where $\tau_0$ is a constant.
The generalization of the definition of the space time effective action $\Sigma_T$ in the situation of tachyon background
was discussed in \cite{tseytlin2}, but in the present context of homogeneous $\beta$-functions, it was argued
in \cite{AKM} that the same definition holds, i.e.
\be\label{SigmaT}
\Sigma_T[g,\phi,T]=-\left(\frac{dZ_T}{d\ln\epsilon}\right)_{\epsilon=1},
\ee
where $Z_T$ is the partition function including the tachyon background~\footnote{We remark at this point that the action 
(\ref{SigmaT}) was dismissed in \cite{tseytlin2} for two reasons. One concerns the fact that this action does not reproduce the 
perturbative Minkowski vacuum, which, however, for us is not a reason, due to the fact that our non perturbative ground 
state is, by construction, not connected to this trivial vacuum configuration. A more serious reason concerns the sign of 
the kinetic term of the tachyon field, which even after passage in the Einstein frame, turns out to be ``ghost-like'', 
i.e. has the wrong sign. However, in our framework, as we shall discuss below, the kinetic term is homogeneous in the 
time field $X^0$. Hence, it can be removed by appropriate field redefinitions~\cite{AKM}, which do not affect the 
string scattering amplitudes. In this sense, the precise form of the tachyon kinetic terms of this action is not 
relevant for our discussion on the existence of dark energy in our configuration.}.
It was then shown that, if the cancellation $2\phi_0+\tau_0=0$ occurs,
the Universe undergoes a de Sitter inflation.
If $2\phi_0+\tau_0\ne 0$, the following scale factor was found
\be\label{scaleT}
a(t)\propto t^{1+\frac{D-2}{2\phi_0+\tau_0}},
\ee
such that the generalization of the condition (\ref{minkowski}) to get a Minkowski Universe is
\be\label{minkowskiT}
D-2+2\phi_0+\tau_0=0.
\ee
In \cite{AKM}, because of the specific $X^0$-dependence of the backgrounds,
the usual tree-level term $-2T$ in the tachyon $\beta$-function, $\beta^T$,
could also be absorbed by a field redefinition, leading to homogeneous $\beta$-functions
$\beta^g_{\mu\nu},\beta^\phi,\beta^T$. In this case,
the steps leading to the equation (\ref{sigma}) would give here the same conclusion: because of the
definition (\ref{SigmaT}), the condition $\dot\Sigma_T=0$ still holds and would lead to a Minkowski Universe.\\
One can consider, though, an out-of-equilibrium situation, where the tachyon
(\ref{tachyon}) is off-shell (but the metric and dilaton are on-shell),
and where the tree-level term $-2T$ is not absorbed in a field redefinition (but
the loop corrections to the tachyon $\beta$-function are homogeneous in time, to all orders in $\alpha'$, since they
depend on the derivatives of the tachyon only, and all contain the same power of $X^0$). In this case,
the space time effective action (\ref{SigmaT}) reads
\bea
\Sigma_T&=&\frac{1}{\kappa_0^2}\int d^Dx\sqrt{|g|}e^{-2\phi-T}\left\{\frac{D-26}{6\alpha'}-\frac{2}{\alpha'}T-
\frac{1}{2}R-\nabla^2\phi-\frac{1}{2}\nabla^2 T+\Delta_T\right\}\nn
&=&\frac{A^{D/2}}{\kappa_0^2}(\alpha')^{D/2-1+\phi_0+\tau_0/2}
\int\frac{d^Dx}{(x^0)^{D+2\phi_0+\tau_0}}\left\{-2\tau_0\ln\left(\frac{x^0}{\sqrt{\alpha'}}\right)+\Theta_T\right\},
\eea
where $\Delta_T$ contains all the higher orders in $\alpha'$, and $\Theta_T$ is a constant independent of $\alpha'$.
The condition $\dot\Sigma_T=0$ gives then
\be\label{eqtheta}
D-2+2\phi_0+\tau_0=\frac{\int d^Dx(x^0)^{-(D+2\phi_0+\tau_0)}}
{\int d^Dx(x^0)^{-(D+2\phi_0+\tau_0)}
\left\{\ln\left(\frac{x^0}{\sqrt{\alpha'}}\right)-\frac{\Theta_T}{2\tau_0}\right\}},
\ee
and if we consider the appropriate field redefinitions in order
to cancel the homogeneous contribution $\Theta_T$, we have then, after simplifying the space volume factors,
\be\label{ratio}
D-2+2\phi_0+\tau_0=\frac{\int dx^0(x^0)^{-(D+2\phi_0+\tau_0)}}
{\int dx^0(x^0)^{-(D+2\phi_0+\tau_0)}\ln(x^0/\sqrt{\alpha'})}\ne 0,
\ee
such that the Minkowski condition (\ref{minkowskiT}) is not satisfied. Instead,
the Universe is either expanding or contracting with the power law
(\ref{scaleT}), depending on the sign of the ratio (\ref{ratio}), where the integrals need to be regularized
in order to avoid a possible initial singularity $x^0=0$. This singularity is indeed present if $D+2\phi_0+\tau_0>1$, what
we now assume and will have to check. Note that
this regularization can't depend on $\alpha'$, since it was assumed when calculating $\dot\Sigma_T$ that the boundaries of the
space time integral are independent of $\alpha'$. If we impose then the integration over $x^0$ to start from a time
$x^0_{init}>\sqrt{\alpha'}$, we can see that the ratio (\ref{ratio}) is positive for $D>2$, which is
consistent with the assumption $D+2\phi_0+\tau_0>1$. The power law (\ref{scaleT}) depends then on the
sign of $2\phi_0+\tau_0$:
\begin{itemize}

\item  A \emph{collapsing} power-law Universe is obtained if
\begin{equation}
2\phi_0 + \tau_0 < 0,
\label{abshor}
\end{equation}
and this situation is ruled out by experimental data;

\item An \emph{accelerating} power-law Universe is obtained if
\begin{equation}
2\phi_0 + \tau_0 > 0
\label{preshor}
\end{equation}
which we interpret as \emph{dark energy} associated with the \emph{off-shell} tachyon field.
This is inflicted by cosmic horizons and hence in such a case perturbative the worldsheet correlation functions
cannot be interpreted as on-shell string scattering amplitudes.
However, in our off-shell case, from a $\sigma$-model point of view, the situation  is well defined, and the background field redefinitions, leading to a scheme guaranteeing the vanishing of the homogeneous terms $\Theta_T$ in (\ref{eqtheta}), are mathematically well defined operations.
\end{itemize}

We next note that, in the case of a more general tachyon potential $V(T)$, 
the linear term in the tachyon $\beta$-function is replaced by 
$-2T\longrightarrow V(T)$. In such a case, 
eq.(\ref{ratio}) becomes
\be\label{ratioV}
D-2+2\phi_0+\tau_0=\tau_0\frac{\int dx^0(x^0)^{-(D+2\phi_0+\tau_0)}V'(T)}
{\int dx^0(x^0)^{-(D+2\phi_0+\tau_0)}V(T)},
\ee
where a prime denotes  derivative with respect to $T$, and we then arrive at a similar conclusion.

A final comment we would like to make concerns \emph{inflation}, or more specifically a de Sitter phase, in this approach, that is a phase characterized by an exponential growth of the scale factor with cosmic time.
As discussed in \cite{AKM}, to arrive at such a phase the presence of \emph{both} tachyon and dilaton fields,
in the configuration (\ref{config}),(\ref{tachyon}), is required with the additional \emph{anti-alignment} condition:
\begin{equation}\label{infl}
2\phi_0 + \tau_0 = 0~.
\end{equation}
In such a case, the Einstein- and $\sigma$-model-frame background fields are the same~\cite{AKM}, and in fact the metric assumes the form of the scale factor of a de Sitter space, $a(t) = e^{H_I t}$, with de Sitter parameter: $H_I = 1/\sqrt{\alpha 'A}$.
The condition (\ref{infl}) spans a very small region (and hence a small probability) in the theory space of strings. Of course, this does not mean that inflation cannot be realized in string theory. It is a matter of initial conditions of our Universe. In the context of colliding branes worlds, for instance, one can imagine a situation in which
the initial tachyonic instabilities have the form (\ref{tachyon}), as a result of the existence of a homogeneous cosmological vacuum energy on the four dimensional brane world, due to interacting strings stretched among the colliding branes, which is almost constant if the collision is adiabatic. In such a case, conformal invariance of strings with their ends attached only to our brane world, which represent matter and radiation, would lead to
vacuum configurations for which the condition (\ref{infl}) is satisfied. This is because, at low energies, a cosmological constant (de Sitter) space would necessarily lead to equations of motion of Einstein/de-Sitter cosmological type with an exponential growth of the scale factor.
Once the Universe enters in such a phase, there is an exponential growth, but there are also further deformations,
in the $\sigma$-model,
which eventually would destroy the anti-alignment condition (\ref{infl}), leading to power law universes (\ref{scaleT}) during and after the exit phase from inflation. The off-shell tachyons in such late era phases would simply describe
effectively off equilibrium effects that still persist at late eras.
Notice that this mechanism does not require the presence of a massive inflaton field, this r\^ole is provided by the \emph{off-shell} tachyon field which leads the inflationary and subsequent growth phases of the Universe.
Eventually the tachyon field will decay away, and as discussed in this work, in such a case the Universe will asymptote to a Minkowski space time, but with a non trivial, logarithmically varying in time, (runaway) dilaton.
This will in turn affect the string coupling $g_s=e^\phi$, which in this scenarios is time dependent.
In the context of four-dimensional models, therefore, this may present a serious problem for phenomenology.
However, such issues are not going to be discussed in this work.

\vspace{0.5cm}

\section{Conclusions\label{sec5}}

In this work we have discussed an approach to deriving an expanding and accelerating or decelerating Universe
by means of \emph{off shell} tachyons, representing instabilities due to cosmically catastrophic events in a string Universe.
The important feature of our approach is the fact that one arrives at such a conclusion without explicit knowledge of 
the target-space string effective action.

Our approach consists of finding field configurations which are basically independent of the magnitude of
$\alpha '$, and hence valid to all orders in a $\sigma$-model-loop expansion for the Weyl anomaly coefficients ($\beta$-functions). 
In our framework, world-sheet conformal invariance is obtained from the homogeneity of these $\beta$-functions in the target 
time $X^0$, by means of appropriate field redefinitions. It should be stressed that in 
the model considered in the present work, such homogeneous terms are spoiled by the tree-level tachyon contributions.
Consequently, as we have seen: {\it (i)} a de Sitter expanding Universe is obtained if the tree level tachyon
contribution is removed by field redefinitions; {\it (ii)} a power-law expanding Universe is obtained if 
this tree level term is left, leading to an off-shell tachyon.

In this sense, the inclusion of such terms leads to non vanishing tachyon $\beta$-functions, and hence to
off-shell tachyon fields. Physically, such off-shell situation may represent the remnant of a cosmically catastrophic event
in the Universe's history, whose (relaxation) effects still last until late eras.

An off-shell tachyon leads in general to a power-law expanding Universe, and naturally
provides a scenario for dark energy.
Under certain conditions among the amplitudes of the dilaton and tachyon configurations
one may also have an inflationary de Sitter phase.

An additional important ingredient of our approach is the possibility to implement this tachyon-induced
dark energy in four dimensions, therefore avoiding the complications arising from compactification.
This latter property, however, poses the challenge to find phenomenologically realistic string theories, with runaway time-dependent dilatons, $\phi (t)$, and hence time-dependent string couplings $g_s =e^\phi$, which necessarily characterize the pertinent non-critical-dimension string configurations.
This is left for future works.

\section*{Acknowledgements}

This work is partially supported by the Royal Society, UK, (J.A.) and by the European Union (N.E.M.)
through the FP6 Marie
Curie Research and Training Network \emph{UniverseNet} (MRTN-CT-2006-035863).


\begin{thebibliography}{99}
\bibitem{hellerman} S.~Hellerman, N.~Kaloper and L.~Susskind,
  JHEP {\bf 0106}, 003 (2001)
  [arXiv:hep-th/0104180];
W.~Fischler, A.~Kashani-Poor, R.~McNees and S.~Paban,
  JHEP {\bf 0107}, 003 (2001)
  [arXiv:hep-th/0104181].



\bibitem{AEM1}
J.~Alexandre, J.~R.~Ellis and N.~E.~Mavromatos,
        JHEP {\bf 0612} (2006) 071
        [arXiv:hep-th/0610072].

\bibitem{aet} J.~Alexandre, N.~E.~Mavromatos and D.~Tanner,
  New J.\ Phys.\  {\bf 10} (2008) 033033
  [arXiv:0708.1154 [hep-th]].


\bibitem{AKM}
J.~Alexandre, A.~Kostouki and N.~E.~Mavromatos,
  JHEP {\bf 0903} (2009) 022
  [arXiv:0811.4607 [hep-th]].

\bibitem{metsaev}
R.~R.~Metsaev and A.~A.~Tseytlin,
  Nucl.\ Phys.\  B {\bf 293} (1987) 385.

\bibitem{AM}
J.~Alexandre and N.~E.~Mavromatos,
  Addendum to JHEP {\bf 0612} (2006) 071
  [arXiv:hep-th/0703171].





\bibitem{gross} D.~J.~Gross and V.~Periwal,
  Phys.\ Rev.\ Lett.\  {\bf 60}, 2105 (1988);

  \emph{ibid.} {\bf 61}, 1517 (1988).



\bibitem{tseytlin1}
A.~A.~Tseytlin,
  Int.\ J.\ Mod.\ Phys.\  A {\bf 4} (1989) 1257.




\bibitem{tseytlin2}
A.~A.~Tseytlin,
  J.\ Math.\ Phys.\  {\bf 42} (2001) 2854
  [arXiv:hep-th/0011033].





\end{thebibliography}
\end{document}